# A bioinformatics system for searching Co-Occurrence based on Co-Operational Formation with Advanced Method (COCOFAM)

Junseok Park
KAIST
335, Gwahakro
Yuseong-gu, Daejeon
Republic of Korea
junseokpark@kaist.ac.kr

Gwangmin Kim
KAIST
335, Gwahakro
Yuseong-gu, Daejeon
Republic of Korea
gwang5386@kaist.ac.kr

Dongjin Jang
KAIST
335, Gwahakro
Yuseong-gu, Daejeon
Republic of Korea
djjang@kaist.ac.kr

Sungji Choo
Yonsei University
College of Medicine
Seoul
Republic of Korea
sjchu1205@gmail.com

Sunghwa Bae
KAIST
335, Gwahakro
Yuseong-gu, Dajeon
Republic of Korea
torch@kaist.ac.kr

Doheon Lee*
KAIST
335, Gwahakro
Yuseong-gu, Daejeon
Republic of Korea
dhlee@kaist.ac.kr

## ABSTRACT
Literature analysis is a key step in obtaining background information in biomedical research. However, it is difficult for researchers to obtain knowledge of their interests in an efficient manner because of the massive amount of the published biomedical literature. Therefore, efficient and systematic search strategies are required, which allow ready access to the substantial amount of literature. In this paper, we propose a novel search system, named "Co-Occurrence based on Co-Operational Formation with Advanced Method" (COCOFAM) which is suitable for the large-scale literature analysis. COCOFAM is based on integrating both Spark for local clusters and a global job scheduler to gather crowdsourced co-occurrence data on global clusters. It will allow users to obtain information of their interests from the substantial amount of literature.

## Categories and Subject Descriptors
H.3.3 [**Information Search and Retrieval**]: Information Search and Retrieval - *Information Filtering*; H.3.4 [**Systems and Software**]: Systems and Software – *Distributed systems*; J.3 [**LIFE AND MEDICAL SCIENCES**]: LIFE AND MEDICAL SCIENCES – *Biology and genetics.*

## General Terms
Algorithms, Management, Documentation, Performance

## Keywords
Co-occurrence, bioinformatics, text-mining, Hadoop, Spark, job-scheduler, crowdsourcing

---

* Doheon Lee is the corresponding author

## 1. INTRODUCTION
Scientific literature plays a key distribution role for significant findings and results from biomedical research, and therefore, literature analysis is an essential step in obtaining background knowledge in biomedical research [1]. However, it is difficult for scientists to isolate knowledge of their interests in an efficient manner, due to the massive amount of the published biomedical literature [2]. In order to support scientists, efficient and systematic search strategies are required, which allow ready access to the substantial amount of literature. For these reasons, several tools have been developed to identify and extract literature of user's interests, such as LitInspector [3], PESCADOR [4], iHOP [5], EBIMed [6], PubGene [7] and PolySearch [8]. However, no tools are available that copes with the computational burden of a large volume of literature as we can introduce.

Recently, with the adoption of both the crowdsourcing paradigm and the Hadoop-based distributed file system, we can efficiently address large-scale computing resource problems and reduce search time [9-12]. Hadoop has many advantages of the parallel computational jobs and distributed jobs by orchestrating the execution of analyzing text data in commodity hardware [13]. Even though Hadoop reduces the search time based on its parallel data flow framework, it has a limitation. The map-reduce algorithm cause disk i/o that is expensive operation, then Spark solved the limitation by its in-memory computations[14]. However, it still has a large-scale computing resource problems which are usually pointed out as a limitation on single cluster environmental Hadoop infrastructure. The globally crowdsourced large pool process in the academic community would solve this limitation. [15, 16]

We suggest a novel search system, named Co-Occurrence based on Co-Operational Formation with Advanced Method (COCOFAM) which allows users to obtain information of their interests from a massive amount of literature by integrating Spark for local clusters and a global job scheduler to gather crowdsourced co-occurrence data on global clusters.

## 2. METHOD

COCOFAM use a number of local clusters as a global cluster. It collects the same tasks from each job of local cluster using a global job scheduler. Then, it distributes the collected tasks to each local cluster for reducing the total number of local tasks on the local job. We also define the result of each task as crowd-sourced [17] co-occurrence data and COCOFAM provides these results with the same purpose. In this paper, *Task* and *Job* are defined as below:

- Task: An individual and single processing refers to the search for a co-occurrence of a paired term.
- Job: Co-occurrences of all paired terms search operation that runs on a local cluster with a specific resource file.

### 2.1 Co-Occurrence Data Set

Co-Occurrence data set in COCOFAM includes not only number of counted co-occurrences but also number of counted each word on each co-occurrence. The data set is required to find statistical information and derive significance of co-occurrences in text resources. We proceed our method based on the definition of co-occurrence data set.

### 2.2 Input File

A user initiates COCOFAM by resource file and paired term list file to execute a job for finding co-occurrences.

#### 2.2.1 Resource File

A resource file is available on the COCOFAM webpage. A user can use the available resource file for finding co-occurrences or, the user can customize a resource file then can provide the file to the COCOFAM webpages as a resource contributor. We used 13,200,786 PubMed abstracts gathered from the 1950 to 2013 as the first resource file for COCOFAM. PubMed abstracts are totally segmented by the GENIA sentence splitter tool [18], resulting in 107,188,823 sentences from all abstracts. The overall resource files have unique identifier (e.g. MD5 hash value) and unique abstract or sentence identifier as a key (e.g. PubMed abstract ID or PubMed sentence of abstract ID).

#### 2.2.2 Paired Term List File

A user should input more than one paired term which is separated by a tab delimiter. Local cluster will load the paired term list file to HDFS, then read each line of the list sequentially.

### 2.3 Task Processing

COCOFAM read paired term list file then start to make a task for finding a co-occurrence.

#### 2.3.1 Term task processing

Loaded paired term list into two columns by delimiter, then duplicated word on each column. The columns are merged to one column and the merged column can also have duplicated word on it. The duplicated words are removed again. Then each word on the column is processed with resource text and get its counts and resource keys which contains the word string. The result will be stored on same row and the total output result of words can be called from next step.

#### 2.3.2 Co-occurrence task processing

One word in paired terms can appear several times and finding each co-occurrence of paired term with ignoring the duplicated attribute is waste of time. Group by more duplicated side column is more efficient method. COCOFAM re-process loaded paired term with duplication information and make a group with each word's resource key information by selecting more duplicated column to save calculation time. Each paired term can find co-occurrence information to intersect each stored resource key information. Resource key is stored as one string value but this information would be convert to table as entity of relational database due to its stored schema information. The restored resource key tables intersect each other then finally calculate co-occurrence of each task. Whole process is described like Figure 1.

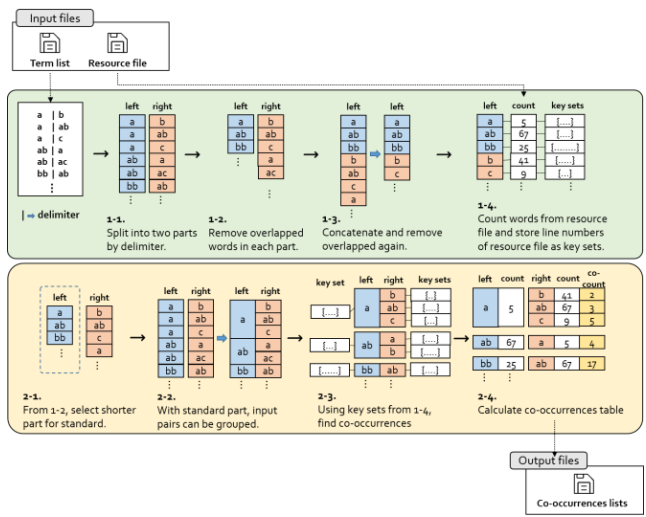

**Figure 1. Task Processing Overview**

### 2.4 Job Processing Controller

Job processing controller is a local program on user's local cluster. The program is run on JVM with Hadoop and Spark libraries and use Parquet file system as a local job processing database and module[19].

YARN resource manager delegates a state-of-the-arts scheduling method, it helps to find a co-occurrence on available local commodity servers by environment configuration information. It automatically recognizes a parallelized task across the local clusters, as it usually has, however, computing resource power limitation on the clusters. Job scheduling modules on global cluster helps it cope with the limitation. The Mapreduce programming model can be extended to global scheduler but we adopt Spark to solve its slowness and lack of a structured data handling method. Accorrdingly we use Spark to be connected with job scheduling modules and allow HDFS or local file to be used as data tables on Spark Dataframe; resource data, paired term list data and task status data [20]. We also use Spark Dataframe to process each task with an SQL-like query language rather than use a raw level key/value computation language on Mapreduce model.

All results of tasks on job processing controller accompany user selected significant values (e.g. Jaccard Index) to show coincidence probability of a co-occurrence for showing its significance.

## 2.5 Job Scheduling Modules

We expand the cluster pool from local to global through the use of job scheduling modules. The modules are designed to reduce the total tasks of each local job since we provide global job pool, crowdsourced co-occurrence results per a job.

### 2.5.1 Job Scheduling Controller

The global cluster is prepared from a unique hash value (e.g. MD5) of a resource file that is analyzed on a local cluster. The job scheduling controller classifies request tasks depending on the following conditions; the existence of crowdsourced data and presence of the same activated task from another local cluster. We used node.js [21] as controller module to handle a task request by http and to return a response by in JSON format [22]. And we constructed job, task management and crowdsourced data list table on MySQL database as it is not only an available structured data management system but it also has DBMS characteristics. [23]

We define crowdsourced data as a usable result of co-occurrence data generated when a local cluster executes a task or co-occurrence search task is done by another local cluster on the global cluster pool.

In first step, the paired term list of a local cluster is shuffled and selected one by one. If the selected term exists in the crowdsourced data, it is directly saved without further task processing. Thus, the number of tasks to be processed can be decreased like

$$T_p = T_t - T_c \quad (1)$$

where $T_p$ is number of tasks to be processed, $T_t$ is number of total tasks, and $T_c$ is number of tasks in crowdsourced data.

If the selected term does not exist in the crowdsourced data, the controller processes a task to one of the following two procedures

(1) If a task does not exist on task management table, it is executed on the local cluster and registered into the task management table with incomplete status. When the task is done, change its status as completed. Then send co-occurrence result to global job scheduler and local controller.

(2) If a task already exists on the task management table on global job scheduler, it is inserted into a pending task list on the local cluster.

After finishing all tasks which was assigned for the local cluster, tasks inserted into a pending task list are processed with the following pseudocode.

```
executePendingTask(String job, InputSet taskSet)
// job : unique hash value of a job
// taskSet : pending state paired term list information in a
job
loop until pending list is not existed
  // task from local Job Scheduler
  task = getOldestTask(taskSet);
  // status from Global Job Scheduler
  taskStatus = getTaskStatus(job, task);
  if taskStatus = 1 then // 1 is complete, 0 is incomplete.
      // Transaction Start
      updateTaskInfo(job,task); // get result and update job
      status on global job scheduler, delete the task on
      pending list of local cluster
      // Transaction End
  else
      executeTask(job,task);
      // the procedure (1) explained above
  end if
```

We use a MySQL transaction characteristic that maintains the integrity and atomic status of a task result processing. Furthermore, all tasks are not duplicated unless a job is disconnected suddenly on the global cluster. Thus, the total number of tasks of a job executed by a local cluster can be decreased like

$$N_{job\ size} = T_t - T_c - T_s * \left(1 - \frac{1}{T_l}\right) + \alpha \quad (2)$$

where $T_t$ is number of total tasks, $T_c$ is number of tasks in crowdsourced data, $T_s$ is number of same tasks, $T_l$ is number of local cluster, and $\alpha$ is number of tasks after releasing pending status.

## 2.6 Job Execution Flow

A co-occurrence search on COCOFAM starts from the local cluster, and then it is expended to the global cluster in order to reduce the total number of tasks on the local cluster. Each job is executed on the local cluster and each task is distributed by the global job scheduler. Figure 2. shows the overall execution procedure of COCOFAM.

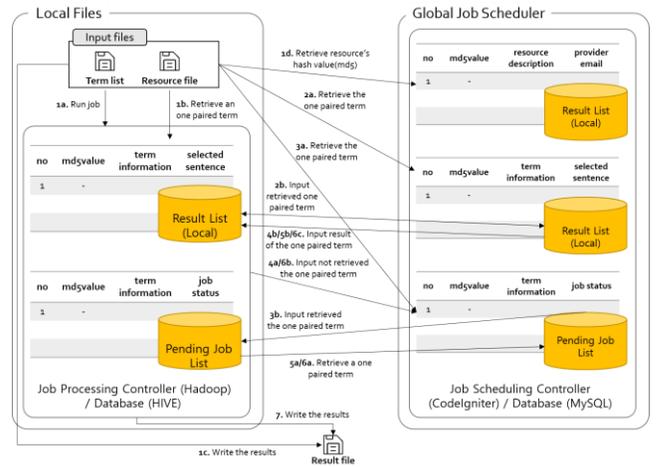

**Figure 2. Overall Job Execution Flow Chart**

## 3. Result

We developed COCOFAM user application on Scala language 2.10.3 with Spark library 1.6.0 and other Hadoop related dependent libraries (e.g. Cloudera CDH 5.7.1-Hadoop 2.6.0 dependencies). If a user wants to use COCOFAM application, firstly set a configuration file then submit COCOFAM application to Spark. A user can check many functions of COCOFAM on the configuration file such as processing mode, capital sensitivity, local database usage, intermediate result storing options. Followings are main functionalities of COCOFAM.

- Standalone Processing Mode : COCOFAM only runs on a user's local server or cluster. It does not connect to the global job scheduler. A user can allocate CPU cores, memories, increase JVM heap size on local server, or deploy the program to their Hadoop cluster with Spark-submit options.

- Co-Operation Processing Mode : This mode includes all functionalities of standard processing mode and provides more functionality with global job scheduler. A user also can enable or disable data transfer option whether send found term information on a resource file to global job scheduler or not. If a user disabled data transfer mode, the user would get limit request size from job scheduler for fair data sharing concept.

We developed global job scheduler on node.js and express framework. The global job scheduler is kind of web application and it provides RestfulAPIs for communicate with Spark executors which deploys COCOFAM application from a user's local cluster. Global job scheduler stores requested data to remote MySQL server and we also designed the schema of MySQL database for providing efficient query executions.

The view page of global job schedulers provides registered resource files, so a user can download the resource file from the view page and also can upload their own resource file for crowdsourcing jobs which the main idea of COCOFAM. Overall processing modes are described on Figure 3.

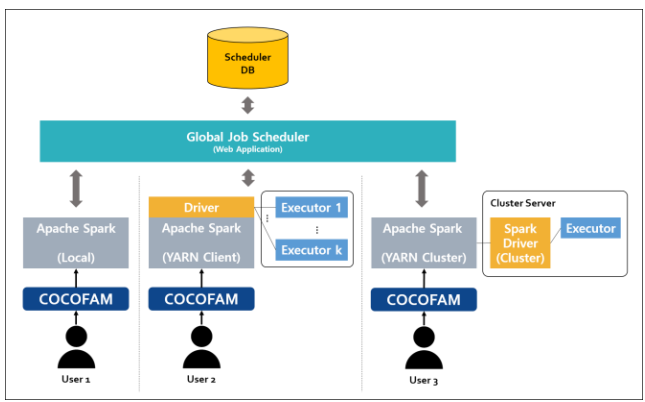

**Figure 3. Several COCOFAM processing modes on overall application architecture**

## 4. Evaluation

We evaluate our result by calculating job processing time on several environments to validate our claims in method section. We organized different evaluation environment for different test scenarios.

**Table 1. Standalone Environment for testing task processing and job processing speed of COCOFAM**

| Hostname | Model | CPU Info | OS | Memory |
|---|---|---|---|---|
| Heart4 | Dell R730 | Xeon E5-2680v3 x2 (24 Cores) | CentOS 6.8 | 256GB |
| chrome000 | Dell R730 | Xeon E5-2680v3 x2 (24 Cores) | CentOS 6.8 | 192GB |

**Table 2. Cluster Environment for testing scalability of COCOFAM**

| Hostname | Model | CPU Info | CDH Role | Memory | Network Speed |
|---|---|---|---|---|---|
| chrome000 | Dell R730 | Xeon E5 2680v3 x2 (24Cores) | name node | 192GB | 6Gbps (bonding) |
| chrome011-013 | Dell R730 | Xeon E5 2630v3 x2 (16Cores) | data node | 64GB | 4Gbps (bonding) |
| Chrome014-016 | HP DL380 | Xeon E5 2630V4 (20Cores) | data node | 64GB | 4Gbps (bonding) |

Firstly, we compared COCOFAM's task processing speed and job processing speed with MySQL. MySQL is common database that is usually used to find text by its SQL query and current web-based tool support many functions in easy manner. Also it is famous RDBMS and well known for its optimized execution plan. We used MySQL 5.6.3 version and also we used phpMyAdmn 5.1.72 for experimental web interface. COCOFAM only used local Spark application and we allocated only 1 CPU core to Spark. Because MySQL also used only 1 CPU core. Although the server specification of MySQL test is slightly better than COCOFAM test server, we suppose that if COCOFAM has better result, it would be more good performance on same server. A resource file for the test was Pubmed Abstracts which size is approximately 4.5GB and input paired term consists of 1 paired term and 10 paired terms. We used MySQL query naturally at first time and applied MySQL query with our task processing method at second time. We found that our task processing method works well and COCOFAM is much faster than MySQL in each test experiments.

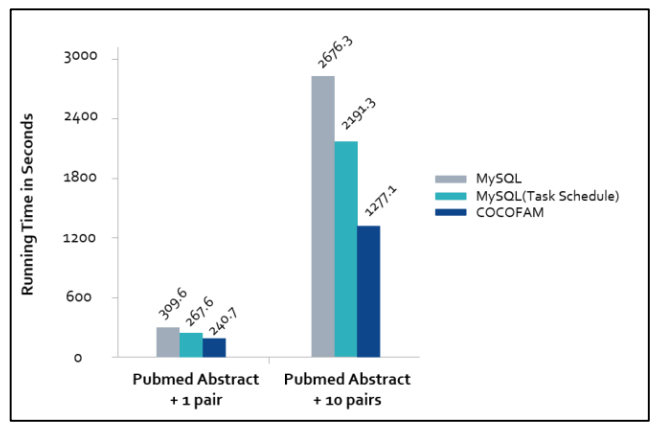

**Figure 4. Speed comparison between MySQL and COCOFAM**

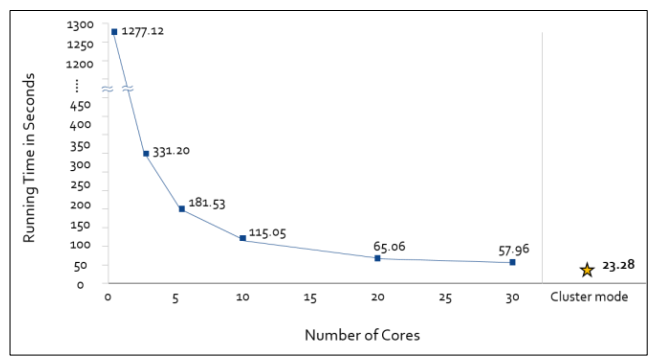

**Figure 5. COCOFAM scalability test results**

Secondly, we tested scalability of COCOFAM. We tested COCOFAM on only single server with different allocated cores and we also tested COCOFAM on our cluster server. The single server is belong to cluster server and cluster servers are installed Cloudera CDH 5.7.1 and Spark 1.6.0. We used HDFS file system to both of tests and we submit COCOFAM to cluster server local mode which stands for single server and yarn-cluster which stands for cluster server. We used yarn-cluster mode with 6 executors, 15 cores each executors, 7.5GB memory to each executors, 16 cores for driver and 46GB memory to driver. We used Pubmed Abstracst and 10 paired terms in every tests. We found that running time of

COCOFAM would be more fast depends on allocated CPU cores and it has great performance on cluster mode.

## 5. DISCUSSION

In this work, our goal is to search co-occurrences from massive biomedical literature with effective means on any Spark with Hadoop clusters. COCOFAM has four attributes to support the goal.

- Scalability: COCOFAM can handle massive biomedical text resources and the resource can be scaled up to the maximum capacity of the total storage size on the local cluster.

- Speed: COCOFAM can significantly increase co-occurrence searching speed depending on the number of nodes and cores on a local cluster.

- Efficiency: The total number of tasks would be reduced through the global cluster as follows

$$1 \leq total\ number\ of\ tasks \leq N_{job\ size} \quad (3)$$

COCOFAM also makes possible the processing of massive jobs with limited computing resource.

- Crowdsourcing: COCOFAM gets any input paired term, then it searches for it on resource file. In the other words, there is no searching space limitation as with the current web-based tool. Thus, it could theoretically service all possible number of co-occurrence results to a biomedical text-mining researcher theoretically.

However, the initialization time is a task to solve for COCOFAM. Initialization time of Spark is notably long. We will support Impala [24], Tajo [25] and Phoenix [26] on the next version to find solution of this problem.

There are also many need to find synonyms when a user find co-occurrence on resource file to avoid duplicated jobs. We have created the synonym ontology for diverse biological terms and their synonym information by integrating publicly-available resources (Table 3) for solving this issue. The collected data will be used when the user finds co-occurrence between two input terms, and the synonym information for each term is provided with the result. However, there were engineering problem with global job scheduler, we will support this feature in next update version of COCOFAM.

**Table 3. Statistics for the Synonym Ontology**

| Entity Type | Reference resources | # of entities | # of total synonyms |
|---|---|---|---|
| Dug | DrugBank | 6825 | 7446 |
| Gene | Entrez, uniprot | 46587 | 107919 |
| Metabolite | HMDB, KEGG, ChEBI | 74879 | 434433 |
| Relation | Text from LMU & public DB | 177 | 100 |
| Organism | MeSH | 3669 | 16431 |
| Disease | MeSH | 4620 | 45856 |
| Organ | MeSH | 1191 | 6832 |
| Tissue | MeSH | 140 | 857 |
| Cell | MeSH | 526 | 3649 |
| Molecular function | Gene Ontology | 3535 | 3416 |
| Cell Compartment | Gene Ontology | 10552 | 26645 |
| Biological process | Gene Ontology | 26294 | 63668 |

## 6. CONCLUSION

We developed COCOFAM to achieve an efficient text-mining technology on biomedical area by co-operation and task distribution. Especially, many users can upload crowdsourced data or execute overlapped task. Those activities make COCOFAM to carry on the purpose. In conclusion, COCOFAM will greatly contribute to develop biomedical text-mining technology. More information on COCOFAM can be found at http://cocofam.kaist.ac.kr/.

## 7. ACKNOWLEDGMENTS

This work was supported by the Bio-Synergy Research Project (NRF-2012M3A9C4048758) of the Ministry of Science, ICT and Future Planning through the National Research Foundation. Out thanks to Jaesup Park for providing resource files to make this tool.

## 8. REFERENCES

[1] Leach, S. M., Tipney, H., Feng, W., Baumgartner, W. A., Kasliwal, P., Schuyler, R. P., Williams, T., Spritz, R. A. and Hunter, L. Biomedical discovery acceleration, with applications to craniofacial development. *PLoS computational biology*, 5, 3 (Mar 2009), e1000215.

[2] Ananiadou, S., Thompson, P., Nawaz, R., McNaught, J. and Kell, D. B. Event-based text mining for biology and functional genomics. *Briefings in functional genomics* (Jun 6 2014).

[3] Frisch, M., Klocke, B., Haltmeier, M. and Frech, K. LitInspector: literature and signal transduction pathway mining in PubMed abstracts. *Nucleic acids research*, 37, Web Server issue (Jul 2009), W135-140.

[4] Barbosa-Silva, A., Fontaine, J.-F., Donnard, E. R., Stussi, F., Ortega, J. M. and Andrade-Navarro, M. A. PESCADOR, a web-based tool to assist text-mining of biointeractions extracted from PubMed queries. *BMC bioinformatics*, 12, 1 (2011), 435.

[5] Hoffmann, R. and Valencia, A. Implementing the iHOP concept for navigation of biomedical literature. *Bioinformatics*, 21 Suppl 2 (Sep 1 2005), ii252-258.

[6] Rebholz-Schuhmann, D., Kirsch, H., Arregui, M., Gaudan, S., Riethoven, M. and Stoehr, P. EBIMed--text crunching to gather facts for proteins from Medline. *Bioinformatics*, 23, 2 (Jan 15 2007), e237-244.

[7] Jenssen, T.-K., Lægreid, A., Komorowski, J. and Hovig, E. A literature network of human genes for high-throughput analysis of gene expression. *Nature genetics*, 28, 1 (2001), 21-28.

[8] Cheng, D., Knox, C., Young, N., Stothard, P., Damaraju, S. and Wishart, D. S. PolySearch: a web-based text mining system for extracting relationships between human diseases, genes, mutations, drugs and metabolites. *Nucleic acids research*, 36, Web Server issue (Jul 1 2008), W399-405.

[9] Anderson, D. P., Korpela, E. and Walton, R. *High-performance task distribution for volunteer computing*. IEEE, City, 2005.

[10] Chervenak, A., Foster, I., Kesselman, C., Salisbury, C. and Tuecke, S. The data grid: Towards an architecture for the distributed management and analysis of large scientific datasets. *Journal of network and computer applications*, 23, 3 (2000), 187-200.

[11] Shvachko, K., Kuang, H., Radia, S. and Chansler, R. *The hadoop distributed file system*. IEEE, City, 2010.

[12] Fuchs, C. Don Tapscott & Anthony D. Williams: Wikinomics: How Mass Collaboration Changes Everything. *International Journal of Communication*, 2 (2008), 11.

[13] Vavilapalli, V. K., Seth, S., Saha, B., Curino, C., O'Malley, O., Radia, S., Reed, B., Baldeschwieler, E., Murthy, A. C., Douglas, C., Agarwal, S., Konar, M., Evans, R., Graves, T., Lowe, J. and Shah, H. Apache Hadoop YARN (2013), 1-16.

[14] Zaharia, M., Chowdhury, M., Das, T., Dave, A., Ma, J., McCauley, M., Franklin, M. J., Shenker, S. and Stoica, I. *Resilient distributed datasets: A fault-tolerant abstraction for in-memory cluster computing*. USENIX Association, City, 2012.

[15] Marx, V. Biology: The big challenges of big data. *Nature*, 498, 7453 (2013), 255-260.

[16] Wang, L., Tao, J., Ranjan, R., Marten, H., Streit, A., Chen, J. and Chen, D. G-Hadoop: MapReduce across distributed data centers for data-intensive computing. *Future Generation Computer Systems*, 29, 3 (2013), 739-750.

[17] Howe, J. The rise of crowdsourcing. *Wired magazine*, 14, 6 (2006), 1-4.

[18] Sætre, R., Yoshida, K., Yakushiji, A., Miyao, Y., Matsubayashi, Y. and Ohta, T. *AKANE system: protein-protein interaction pairs in BioCreAtIvE2 challenge, PPI-IPS subtask*. City, 2007.

[19] Floratou, A., Minhas, U. F., Fatma, #214 and zcan SQL-on-Hadoop: full circle back to shared-nothing database architectures. *Proc. VLDB Endow.*, 7, 12 (2014), 1295-1306.

[20] Armbrust, M., Xin, R. S., Lian, C., Huai, Y., Liu, D., Bradley, J. K., Meng, X., Kaftan, T., Franklin, M. J. and Ghodsi, A. *Spark sql: Relational data processing in spark*. ACM, City, 2015.

[21] Tilkov, S. and Vinoski, S. Node. js: Using JavaScript to build high-performance network programs. *IEEE Internet Computing*, 14, 6 (2010), 80.

[22] Crockford, D. The application/json media type for javascript object notation (json) (2006).

[23] McCarthy, D. and Dayal, U. *The architecture of an active database management system*. ACM, City, 1989.

[24] Kornacker, M. and Erickson, J. Cloudera Impala: Real Time Queries in Apache Hadoop, For Real. *ht tp://blog. cloudera. com/blog/2012/10/cloudera-impala-real-time-queries-in-apache-hadoop-for-real* (2012).

[25] Choi, H., Son, J., Yang, H., Ryu, H., Lim, B., Kim, S. and Chung, Y. D. *Tajo: A distributed data warehouse system on large clusters*. IEEE, City, 2013.

[26] *Apache Phoenix*. City, 2015.